\documentstyle[epsfig,longtable]{aipproc}
\renewcommand

\begin{document}

\title{
\vspace{-2.5cm}
\flushleft{\normalsize DESY 97-116} \hfill\\
\vspace{-0.15cm}
{\normalsize HUB-EP-97/37} \hfill\\
\vspace{-0.15cm}
{\normalsize June 1997} \hfill\\
\vspace{0.5cm}
\centering{Hadron Structure Functions from} \\[0.3em]
\centering{Lattice QCD -- 1997}}

\author{C. Best$^a$, M. G\"ockeler$^b$, R. Horsley$^c$, L. Mankiewicz$^d$, 
H. Perlt$^e$, P. Rakow$^f$, A. Sch\"afer$^b$,  
G.~Schierholz$^{f,g}$~\footnote{Minireview given by G. Schierholz at 
{\it DIS97}, Chicago, April 1997}, A. Schiller$^e$,
S.~Schramm$^h$ and P. Stephenson$^f$}
\address{$^a$ Institut f\"ur Theoretische Physik, Johann Wolfgang 
Goethe-Universit\"at, D-60054~Frankfurt \\
$^b$ Institut f\"ur Theoretische Physik, Universit\"at Regensburg,
D-93040 Regensburg \\
$^c$ Institut f\"ur Physik, Humboldt-Universit\"at, D-10115 Berlin \\
$^d$ Institut f\"ur Theoretische Physik, Technische Universit\"at M\"unchen,
D-85747 Garching \\
$^e$ Institut f\"ur Theoretische Physik, Universit\"at Leipzig, 
D-04109 Leipzig \\
$^f$ Deutsches Elektronen-Synchrotron DESY, Institut f\"ur Hochenergiephysik 
und HLRZ, D-15735 Zeuthen \\
$^g$ Deutsches Elektronen-Synchrotron DESY, D-22603 Hamburgy \\
$^h$ Gesellschaft f\"ur Schwerionenforschung GSI, D-64220 Darmstadt}

\maketitle

\begin{abstract}
We review the status of lattice calculations of the deep-inelastic 
structure functions of the nucleon. In addition, we present some results
on the pion and rho structure functions.
\end{abstract}

\section*{Introduction}

The calculation of the deep-inelastic structure functions of hadrons
from {\it QCD} is basically a non-perturbative problem. 
Perturbation theory can only describe the evolution of the structure 
functions from one scale to the other, provided one is at high enough scales.
The most promising tool to solve this problem is lattice gauge theory.

Three years ago we have initiated a program to compute the lower moments 
of the structure functions of the nucleon on the lattice. First results for 
the unpolarized structure functions $F_1$, $F_2$ and the polarized structure
functions $g_1$, $g_2$ are now available~\cite{1,2,3}. It has been found
that with the help of dedicated computers one is able to compute the lower
moments of all four structure functions with satisfactory precision -- at 
least for the non-singlet leading twist contributions 
which we have addressed so far. It is needless to say that all these 
calculations have been
performed in the quenched approximation, where the effect of internal quark
loops has been neglected. Three or four 
moments are generally sufficient to 
determine the individual quark distribution 
functions~\cite{mw}.

Recently, we have extended our calculations to the structure functions of
the pion and the rho~\cite{best}. The pion structure function is being
measured at {\it HERA}, and data should soon become available. The structure
functions of the rho are interesting because they give qualitatively new
information about quark binding effects.

Lattice calculations are subject to systematic errors arising from the
finite lattice spacing, the finite volume, the extrapolation to the
chiral limit and, of course, quenching. Before we can trust our results
entirely, it is important to examine these effects carefully. This will be 
a major task in the future.

The aim of this talk is twofold. In the first part  
we will present and discuss our results obtained over the last two years. 
This will be mainly about the nucleon structure functions and briefly about
the structure functions of the pion and the rho.
In the second part of the talk we will 
report on first attempts to reduce discretization errors.

\section*{Basics}
 
The theoretical basis of the calculation is the operator product expansion. 
For large $Q^2$ the moments of the nucleon structure functions are given by
\begin{eqnarray}
2 \int_0^1 dx x^{n-1} F_1(x,Q^2) 
&=& \sum_{f=u,d,g} c^{(f)}_{1,n}(\mu^2/Q^2,g(\mu))\: v_n^{(f)}(\mu),
\nonumber \\ 
\int_0^1 dx x^{n-2} F_2(x,Q^2) 
&=& \sum_{f=u,d,g} c^{(f)}_{2,n}(\mu^2/Q^2,g(\mu))\: v_n^{(f)}(\mu), 
\nonumber \\
2\int_0^1 dx x^n g_1(x,Q^2) 
  &=& \frac{1}{2} \sum_{f=u,d,g} e^{(f)}_{1,n}(\mu^2/Q^2,g(\mu))
\: a_n^{(f)}(\mu),  
 \\
2\int_0^1 dx x^n g_2(x,Q^2) 
  &=& \frac{1}{2}\frac{n}{n+1} \sum_{f=u,d,g} [e^{(f)}_{2,n}(\mu^2/Q^2,g(\mu))
\: d_n^{(f)}(\mu) \nonumber \\
  &-& e^{(f)}_{1,n}(\mu^2/Q^2,g(\mu))\: a_n^{(f)}(\mu)], \nonumber
\label{in1}
\end{eqnarray}
where $c_1$, $c_2$ and $e_1$, $e_2$ are the Wilson coefficients, and $v_n$,
$a_n$ and $d_n$ are forward nucleon matrix elements of certain local 
operators ${\cal O}$. For details see (e.g.) ref.~\cite{1}.
In parton model language
\begin{eqnarray}
v_{n+1}^{(f)} &=& \langle x^{n} \rangle^{(f)} = \int_0^1 \mbox{d}x x^n \frac{1}{2}
[q_\uparrow^{(f)}(x) + q_\downarrow^{(f)}(x)], \; f=u, d,
\label{in4} \\
v_{n+1}^{(g)} &=& \langle x^{n} \rangle^{(g)} = \int_0^1 \mbox{d}x x^n g(x),
\label{in4a} 
\end{eqnarray}
and 
\begin{equation}
a_n^{(f)} = 2 \Delta^{(n)} q^{(f)} = 2 \int_0^1 \mbox{d}x x^n \frac{1}{2}
[q_\uparrow^{(f)}(x) - q_\downarrow^{(f)}(x)], \; f=u, d
\label{in5}
\end{equation}
with $\Delta^{(0)} q^{(f)} \equiv \Delta q^{(f)}$, and $q_\uparrow(x)$,
$q_\downarrow(x)$ being the quark distribution functions with spin $\uparrow$,
$\downarrow$ with respect to the direction of motion. In 
eqs.~(\ref{in4}), (\ref{in5}) we have omitted the sea
quark contributions because we are working in the quenched approximation.
The matrix elements $d_n$ have twist three and so have no parton model
interpretation.

For a spin-zero target like the pion one has 
$q_\uparrow(x) = q_\downarrow(x)$.
For a spin-one target like the rho the quark-spin distribution will depend
on the spin projection $m$ of the particle. Writing $q^m(x) = 
1/2\, [q^m_\uparrow(x) + q^m_\downarrow(x)]$, one finds the new structure 
function~\cite{jaffe}
\begin{equation}
b_1(x) = q^0(x) - q^1(x).
\end{equation}

On the lattice one computes
bare operators ${\cal O}(a)$, where $a$ is the lattice spacing. These
operators are in general divergent and must be 
renormalized. One may define finite operators renormalized at the finite 
scale $\mu$ by
\begin{equation}
{\cal O}(\mu) = Z_{\cal O}(a\mu,g(a)) {\cal O}(a).
\end{equation}
The renormalization condition here must match the renormalization condition 
used to calculate the Wilson coefficients, so that the $\mu$-dependence drops
out of the product of Wilson coefficients and $Z_{\cal O}$'s. 
In principle the renormalization constants can be calculated in
perturbation theory. The problem is only that lattice perturbation 
theory converges very slowly~\cite{lm}.

\section*{Nucleon Structure Functions}

The lattice calculation of structure functions divides into two parts. 
Part one is the 
calculation of the hadron matrix elements. Part two is the 
determination of the renormalization constants $Z_{\cal O}$. Altogether we
have considered $O(20)$ different operators so far. On the hypercubic lattice
one is limited to operators with spin $\leq 4$. In the notation of 
eqs.~(\ref{in4})-(\ref{in5}) this means $n \leq 3$.

For nearly all of the operators the renormalization constants have been
computed perturbatively to one loop order~\cite{c,4,5}. A non-perturbative
calculation has only been started recently~\cite{m,mel}. In the few cases 
where we can compare perturbative and non-perturbative results we see
differences of not more than 5\% if boosted perturbation theory~\cite{lm}
is applied.

The numerical calculations of the matrix elements have so far been done at 
one value of the coupling, 
$\beta \equiv 6/g^2 = 6.0$. 
At this coupling the lattice spacing is $a \approx 0.1 \, \mbox{fm}$.
We did calculations on two volumes, $16^3 \times 32$ and $24^3 \times 32$,
and for several quark masses ranging from $\,\approx \, 30$ to $200 \, \mbox{MeV}$. \hfill
It is important to 

\clearpage

\begin{table}
\caption{Results for the unpolarized structure functions in the chiral
limit. The lattice numbers are compared with the 
{\it CTEQ3M}~\protect\cite{cteq}
fit of the valence quark distribution functions. For $\langle x \rangle$,
which has been calculated using different procedures, we have averaged
over the results in ref.~\protect\cite{2}.}
\begin{tabular}{lll} 
\multicolumn{1}{c}{Moment}  & \multicolumn{1}{c}{Lattice} & 
\multicolumn{1}{c}{Experiment} \\
  & \multicolumn{1}{c}{(quenched, $\mu^2 \approx 5\, GeV^2$)} & 
\multicolumn{1}{c}{($\mu^2 = 4\, GeV^2$)} \\[0.3em] \tableline 
$\langle x \rangle^{(u)}$ & \hspace*{1.00cm} 0.410(34) & \hspace*{0.55cm} 
0.284 \hspace*{0.1cm}   \\
$\langle x \rangle^{(d)}$ & \hspace*{1.00cm} 0.180(16) & \hspace*{0.55cm} 
0.102 \hspace*{0.1cm} \\[0.3em] 
$\langle x \rangle^{(u)}$-$\langle x \rangle^{(d)}$ & 
\hspace*{1.00cm} 0.230(38) & \hspace*{0.55cm} 
0.182  \\[0.3em] 
$\langle x^2 \rangle^{(u)}$ & \hspace*{1.00cm} 0.108(16) &
\hspace*{0.55cm} 0.083 \hspace*{0.1cm}  \\ 
$\langle x^2 \rangle^{(d)}$ & \hspace*{1.00cm} 0.036(8) &
\hspace*{0.55cm} 0.025 \hspace*{0.1cm}  
\\[0.3em] 
$\langle x^3 \rangle^{(u)}$ & \hspace*{1.00cm} 0.020(10) &
\hspace*{0.55cm} 0.032 \hspace*{0.1cm}  \\ 
$\langle x^3 \rangle^{(d)}$ & \hspace*{1.00cm} 0.000(6) &
\hspace*{0.55cm} 0.008 \hspace*{0.1cm} 
\\[0.3em] 
$\langle x \rangle^{(g)}$ & \hspace*{1.00cm} 0.53(23) &
\hspace*{0.55cm} 0.441 \\[-0.2em]
 &  & 
\end{tabular}
\end{table}

\begin{table}
\caption{Results for the polarized structure functions in the chiral limit.
The lattice numbers are compared with experiment~\protect\cite{exp,pretz} 
and the 
{\it LO}~\protect\cite{gs} fit to the valence quark distribution functions. 
The latter
are referred to as {\it valence}. The data for $\Delta^{(1)} u$ and 
$\Delta^{(1)} d$ refer to $\mu^2 = 10\,\mbox{GeV}^2$.}
\begin{tabular}{lll} 
\multicolumn{1}{c}{Moment}  & \multicolumn{1}{c}{Lattice} & 
\multicolumn{1}{c}{Experiment}  \\
  & \multicolumn{1}{c}{(quenched, $\mu^2 \approx 5\, GeV^2$)} & 
\multicolumn{1}{c}{($\mu^2 = 3-5\, GeV^2$)}  \\[0.3em] 
\tableline
$\Delta u$ & \hspace*{1.0cm} 0.841(52) & \hspace*{1.5cm} 0.823 \hspace*{0.1cm} \hspace*{0.4cm} $valence$ 
\\[0.3em] 
$\Delta d$ & \hspace*{0.83cm} --0.245(15) & \hspace*{1.33cm} --0.303 \hspace*{0.1cm} \hspace*{0.4cm} $valence$ 
\\[0.3em] 
$g_A$ & \hspace*{1.0cm} 1.086(67) & \hspace*{1.5cm} 1.26 
\\[0.3em] 
$\int_0^1 \mbox{d}x (g_1^{p} - g_1^{n})$ & \hspace*{1.0cm} 0.176(14) & \hspace*{1.5cm} 0.163(10)(16)  
\\[0.3em] 
$\Delta^{(1)} u$ & \hspace*{1.0cm} 0.198(8) & \hspace*{1.5cm} 0.169(18)(12) 
\\[0.3em] 
$\Delta^{(1)} d$ & \hspace*{0.83cm} --0.048(3) & \hspace*{1.33cm} 
--0.055(27)(11) 
\\[0.3em] 
$\int_0^1 \mbox{d}x x^2 g^{(p)}_1$ & \hspace*{1.0cm} 0.0150(32) & \hspace*{1.5cm} 0.012(1) 
\\[0.3em] 
$\int_0^1 \mbox{d}x x^2 g^{(n)}_1$ & \hspace*{0.83cm} --0.0012(20) & \hspace*{1.33cm} --0.004(3)  
\\[0.3em] 
$\int_0^1 \mbox{d}x x^2 g^{(p)}_2$ & \hspace*{0.83cm} --0.0261(38) & \hspace*{1.33cm} --0.006(2)  
\\[0.3em] 
$\int_0^1 \mbox{d}x x^2 g^{(n)}_2$ & \hspace*{0.83cm} --0.0004(22) & \hspace*{1.5cm} 0.005(8) \\[-0.2em]
 & &  
\end{tabular}
\end{table}

\clearpage
\begin{figure}[t!] 
\centerline{\epsfig{file=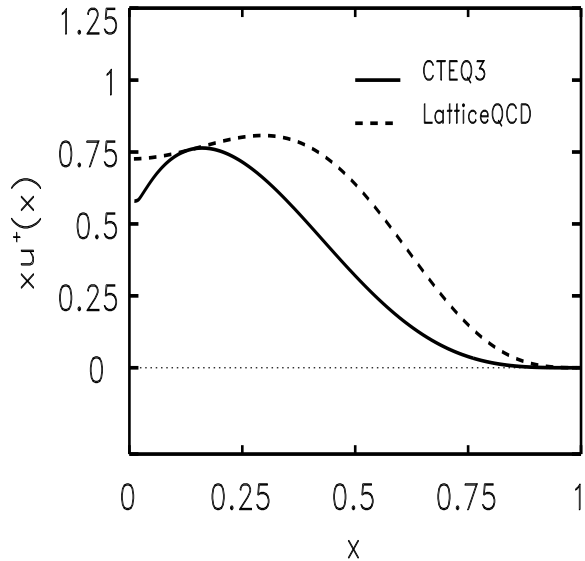,height=3.5in,width=3.5in}}
\vspace{10pt}
\caption{The $u$-quark distribution function extracted from the lattice
data compared with the experimental valence quark distribution function.
The figure is taken from ref.~\protect\cite{mw}.}
\label{fig1}
\end{figure}

\noindent
have data at many quark masses 
to do a
reliable extrapolation to the 
chiral limit. The calculation of nucleon matrix elements, 
in particular 
of higher derivative operators, requires high statistics. Our 
calculations on the
$16^3 \times 32$ lattice involved $O(1000)$ gauge field configurations,
and $O(100)$ configurations on the $24^3 \times 32$ lattice.
The calculations on the large volume were nearly as expensive in computer 
time as the high statistics calculations on the smaller volume as we went to
smaller quark masses. For the calculation of the gluon distribution function
we even used $O(5000)$ configurations. By comparing the results on the two
volumes, we find that our calculations on the $16^3 \times 32$ lattice do
not suffer from finite size effects. All results 
in this and the next section are for Wilson fermions.

Our results for the unpolarized structure functions are listed in table 1, and
in table 2 we give the results for the polarized structure functions. 
Recent work on $\Delta q$ by two other groups~\footnote{These groups also 
compute the sea quark contribution which one might object to in a quenched
calculation.}~\cite{j,l} give similar numbers to ours. We compare the lattice
results directly with experiment, where possible, and otherwise with the
phenomenological valence quark distribution functions, namely 
{\it CTEQ3M}~\cite{cteq} for the unpolarized structure functions and 
{\it LO}~\cite{gs} for the polarized structure functions.

Let us now discuss the results in some detail. We begin with the unpolarized
structure functions in table 1. We see that the lattice prediction for the 
lowest moment, $\langle x \rangle$, is significantly larger than the lowest
moment of the phenomenological valence quark distribution function, both
for the $u$ and the $d$ quark. When one goes to the higher moments the agreement
improves.

What does this mean for the structure functions themselves? In \cite{mw} 
the moments from table 1 have been converted to `real' structure
functions. In fig. 1 the result is shown for the $u$-quark distribution
function. This is compared with the phenomenological curve. We see that the 
`lattice' structure function is quite different in shape from the
experimental one. The difference shows mainly at $x$ values around $1/2$.
We do not expect that the missing internal quark loops will have an effect
at such large values of $x$. If the effect is real, it could mean that the
experimental structure functions contain substantial higher-twist 
contributions. One would expect higher-twist contributions to show up
predominantly at large $x$.

The lowest moment of the gluon distribution function, 
$\langle x \rangle^{(g)}$, turns out to be consistent with experiment.
The relatively large error has to do with the fact that the calculation
requires a delicate subtraction between two terms similar in 
magnitude~\cite{3}. Quark and gluon contributions to $\langle x \rangle$
must add up to one, as a result of energy-momentum conservation. We 
obtain
\begin{equation}
\langle x \rangle^{(u)} + \langle x \rangle^{(d)} + \langle x \rangle^{(g)} 
= 1.12(23).
\end{equation}

Let us now turn to the polarized structure functions in table 2. Here we
have a lot of experimental data to compare with, so that we do not have to 
rely only on the phenomenological distribution functions. A recent, and 
yet unpublished, piece of information comes from the {\it SMC}
measurement of the charge asymmetry of fast pions in polarized muon-nucleon
collisions~\cite{pretz}. In the asymmetry the effect of sea quarks drops out,
and one is lead directly to the moments of the valence quark distribution
functions
\begin{equation}
\Delta^{(1)} u = \frac{1}{2} a_1^{(u)}, \; \Delta^{(1)} d = 
\frac{1}{2} a_1^{(d)}.
\end{equation}
The lattice calculation of $\Delta^{(1)} u$
and $\Delta^{(1)} d$ is reported here for the first time~\cite{m&us}.
For the higher moments sea quark effects should not play any significant
role any more because they are restricted to the region of small $x$.

With two exceptions, the moments of the polarized structure functions agree 
surprisingly well with experiment. 
One exception is the axial vector coupling constant 
$g_A = \Delta u - \Delta d$. Experimentally, this is one of the best known 
quantities, and the sea quark contribution is expected to cancel out. The
lattice result turns out to be quite a bit lower than the experimental
value. The other exception is the twist-three matrix element $d_2$,
contributing to the second moment of the structure function $g_2$. 
Comparing existing data on $g_1$ and $g_2$, one is
lead to the result $d_2 \approx 0$, both for the $u$ and the $d$ quark. We find
that $d_2 \approx - a_2$ in the chiral limit. In the heavy quark limit, on
the other hand, $d_2$ vanishes as one would expect. We have no explanation
for this discrepancy. We do not think that it can be explained by finite
cut-off effects and the effect of quenching.

\section*{Pion and Rho Structure Functions}

The pion and rho structure functions have been computed from $O(500)$ 
gauge field configurations on the $16^3 \times 32$ lattice. As before,
the coupling was taken to be $\beta = 6.0$.

For the pion we obtain in the chiral limit
\begin{equation}
\langle x \rangle = 0.274(13), \, \langle x^2 \rangle = 0.107(35), \,
\langle x^3 \rangle = 0.048(20).
\end{equation}
If we compare this result with indirect information from the Drell-Yan cross
section~\cite{DY}, we find the same picture as for the nucleon: 
$\langle x \rangle$ comes out to be significantly larger than the 
phenomenological value, while $\langle x^2 \rangle$
and $\langle x^3 \rangle$ are consistent with the data.

The unpolarized rho structure function looks very similar to the pion
structure function. For the moment $\Delta q$ (the counterpart of 
$\Delta^{(0)} q$ in eq.~(\ref{in5}) with $m = 1$ instead of $1/2$) of the 
polarized structure function we find $0.59(5)$ indicating that the valence
quarks carry 60\% of the spin of the rho. This is about the same fraction
as one finds for the nucleon. The lowest moment of the polarized structure function
$b_1$ turns out to be positive and unexpectedly large, albeit with large
statistical errors. A possible interpretation of this is that the valence
quarks have a significant orbital angular momentum.

\section*{Recent Developments}

Wilson fermions (which we are using here) give rise to systematic errors of 
$O(a)$, while the standard gluonic action induces errors of 
$O(a^2)$ only. Therefore it is the fermionic action which is most in need
of attention. From hadron mass calculations we know that discretization
errors can have a significant effect on the results at present
values of the coupling~\cite{mass}. 

Since it is very expensive to reduce cut-off effects by reducing $a$, a better
way of reducing them is by improving the action. A systematic improvement
program reducing the cut-off errors order by order in $a$ has been 
proposed by Symanzik~\cite{Sy} and was further developed in~\cite{L&W}.
By adding the (irrelevant) term
\begin{equation}
c \sum_x \bar{\psi} \sigma_{\mu\nu} F_{\mu\nu} \psi
\end{equation}
to the (Wilson) action~\cite{sw} and choosing the coefficient $c$ 
appropriately~\cite{alpha}, one can reduce the cut-off errors from $O(a)$
to $O(a^2)$, provided the operators are improved as well.

\clearpage
\begin{figure}[b!] 
\vspace*{-1.2cm}
\centerline{\epsfig{file=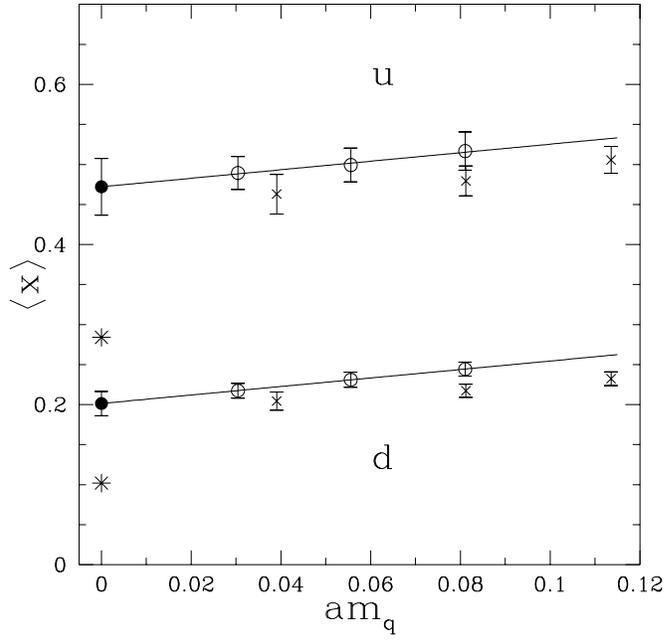,height=3.5in,width=3.5in}}
\vspace{10pt}
\caption{The moment $\langle x \rangle$ for $u$ and $d$ quark distributions 
as a function of the quark mass at $\beta = 6.0$. We compare results using 
the Wilson 
action ($\circ$) with results from the improved action ($\times$). The lines 
are a linear extrapolation of the Wilson action data to the chiral limit. The 
extrapolated values ($\bullet$) are compared with the {\it CTEQ} results
($+\hspace{-0.27cm}\times$).}
\end{figure}

\begin{figure}[b!] 
\centerline{\epsfig{file=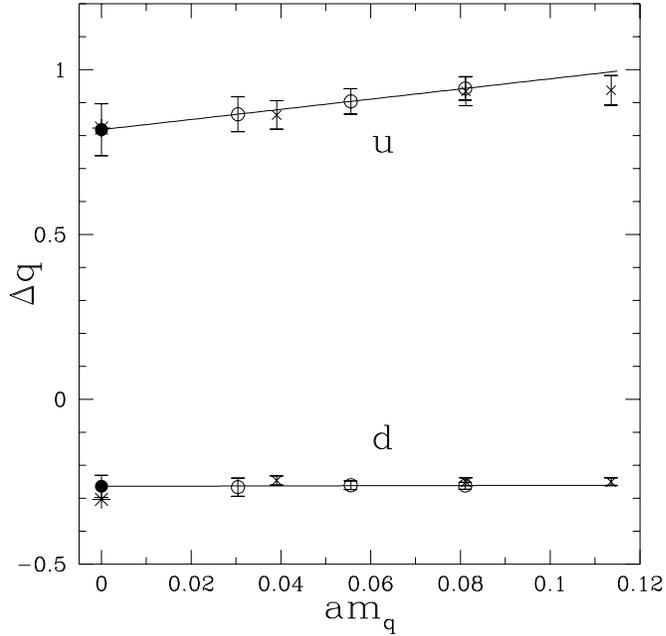,height=3.5in,width=3.5in}}
\vspace{10pt}
\caption{The same as fig.~1 but for $\Delta q$. Here the extrapolated values 
are compared with the phenomenological distribution {\it LO} of 
ref.~\protect\cite{gs}.}
\end{figure}

In figs. 2 and 3 we show first results of the $O(a)$ improved theory
for the moments\footnote{For details
see~\cite{usimp}. For the experts we like to mention that for $\Delta q$ the 
improved operators and the renormalization constants were taken 
from~\cite{alpha,alpha2}, while for $\langle x \rangle$ we used improved 
operators and renormalization constants suggested by
tadpole improved perturbation theory~\cite{cus}.} 
$\langle x \rangle$ and $\Delta q$.
The calculations are done on the $16^3 \times 32$ lattice at $\beta = 6.0$.
We compare the improved results with the unimproved numbers. 
To facilitate the comparison, we have only looked at one operator on one 
lattice.
In the case of 
$\Delta q$ we see no difference between the two results at any value of the 
(bare) quark mass. In the case of the moment $\langle x \rangle$ we see only 
a small effect, which is however not statistically significant. For both
moments the quark mass dependence is rather weak and unspectacular, so that
a linear extrapolation to the chiral limit is justified. 

We may conclude that our numbers for $\langle x \rangle$ 
and $\Delta q$ are not far from the continuum values (of the quenched
theory). This means that we have a real problem with the 
unpolarized structure functions. It is very likely that the solution lies in
unexpectedly large higher-twist contributions at our values of $Q^2$.
The situation of the axial vector coupling constant $g_A$ has also not 
changed. One should, however, keep in mind that the naive parton model
identification of $g_A$ with $\Delta u - \Delta d$ might fail.

Whether discretization errors are also small for the other moments remains 
to be seen.

\vspace{-0.3cm}

\section*{Conclusions}

We have reviewed the current status of lattice structure function calculations.
The results so far are encouraging. But it is clear that much more remains
to be done. We have only begun to improve the calculation by systematically
removing discretization errors, and this will certainly occupy us throughout
the next year.

A topic which is of upcoming interest~\cite{levine} is power corrections (see
also above). This includes higher twist effects on the one hand and 
renormalon contributions on the other. We hope to be able to present first 
results on this topic at next year's workshop.

\end{document}